\documentclass[aps,showpacs,prb,reprint,superscriptaddress,longbibliography]{revtex4-1}

\usepackage{graphicx,bm,times}
\usepackage{color}
\usepackage{amsmath,amssymb}
\usepackage{setspace}
\usepackage{enumerate}
\usepackage{subcaption}
\usepackage{mathtools}
\usepackage[colorlinks=true,linkcolor=blue,citecolor=blue,urlcolor=blue]{hyperref}
\usepackage[font=small]{caption}
\usepackage{multirow}
\usepackage{braket}

\begin{document}

\title{Using forces to accelerate first-principles anharmonic
  vibrational calculations}

\author{Joseph C.\ A.\ Prentice} 
\affiliation{TCM Group, Cavendish  Laboratory, University of Cambridge, 
  J.\ J.\ Thomson Avenue, Cambridge, CB3 0HE, United Kingdom}

\author{R.\ J.\ Needs}
\affiliation{TCM Group, Cavendish  Laboratory, University of Cambridge, 
  J.\ J.\ Thomson Avenue, Cambridge, CB3 0HE, United Kingdom}

\date{\today}

\begin{abstract}
  
  High-level vibrational calculations have been used to investigate
  anharmonicity in a wide variety of materials using
  density-functional-theory (DFT) methods.  We have developed a new
  and efficient approach for describing strongly-anharmonic systems
  using a vibrational self-consistent-field (VSCF) method. By far the
  most computationally expensive part of the calculations is the
  mapping of an accurate Born-Oppenheimer (BO) energy surface within
  the region of interest. Here we present an improved method which
  reduces the computational cost of the mapping. In this approach we
  use data from a set of energy calculations for different vibrational
  distortions of the materials and the corresponding forces on the
  atoms. Results using both energies and forces are presented for the
  test cases of the hydrogen molecule, solid hydrogen under high
  pressure including mapping of two-dimensional subspaces of the BO
  surface, and the \textit{bcc} phases of the metals Li and Zr. The
  use of forces data speeds up the anharmonic calculations by up to
  40\%.
\end{abstract}

\maketitle

\section{Introduction}

Developing an accurate theoretical model for crystalline solids
requires going beyond the static lattice model and including the
effects of atomic vibrations. Usually this is achieved within the
harmonic\cite{wallace_thermodynamics_1972,maradudin_theory_1971} or
quasiharmonic approximations - the latter includes thermal expansion
by making the phonon frequencies volume
dependent\cite{pavone_ab_1993,baroni_phonons_2001}. In many materials
and situations these approximations work very well. However, they may
be inaccurate if the vibrational amplitudes are large. Large
amplitudes may occur, for example, in materials that contain light
elements, materials close to a structural instability, or at high
temperatures. Under such conditions it may be necessary to include
anharmonic vibrations in the model. The mean-field
vibrational-self-consistent-field (VSCF) method proposed by Monserrat,
Drummond and Needs\cite{monserrat_anharmonic_2013} provides a
sophisticated approach for computing the anharmonic vibrational
wavefunction and energy of crystalline solids from first
principles. This approach has been applied successfully in several
systems\cite{monserrat_electron-phonon_2014,monserrat_temperature_2014,drummond_quantum_2015,engel_vibrational_2016,prentice_first-principles_2017}.

Within the Born-Oppenheimer (BO) approximation\cite{born_zur_1927} the
electronic and nuclear motion can be separated, giving two coupled
Schr\"{o}dinger equations.  Application of the VSCF method consists of
two main tasks - mapping of the BO energy surface (the ``energy
landscape'') that the nuclei move through, and self-consistently
solving the VSCF equations for the anharmonic vibrational wavefunction
and energy. Of these two tasks, the former is by far the most
expensive, as it is usually accomplished by performing a large number
of single-point density-functional-theory (DFT) calculations and
fitting a functional form to the calculated energies. To ensure that
the fitted form of the BO surface is accurate, it is necessary to
converge the solution of the VSCF equations with respect to the number
of mapping points used. Reducing the required number of mapping points
reduces the computational cost of mapping the BO surface. The
development of a method for increasing the accuracy of the fit for a
given number of mapping points would help in reaching this goal.

Here we propose a scheme for improving the accuracy of the fitted
functional form of the BO surface by making use of the force data
generated by the DFT calculations in addition to the total electronic
energy. Accurate forces are readily available within plane-wave basis
DFT\cite{payne_iterative_1992}. The gradient of the BO surface at each
mapping point can be obtained from the forces, which provides
additional data that can be used in the fitting process and therefore
provides a better fit with negligible additional cost. We have
investigated the effectiveness and cost of the VSCF method with forces
(VSCF+f) and without forces (VSCF) for several systems in which
anharmonicity is expected to play an important role - the hydrogen
molecule, solid hydrogen at 100~GPa, and the \textit{bcc} phases of
lithium and zirconium, which exhibit soft modes. The effect of
including both 1-D and 2-D subspaces of the BO surface in the
calculations for solid hydrogen, corresponding to including coupling
between anharmonic phonons, is also considered.

As the lightest element, hydrogen is an excellent test case for the
VSCF+f method, as its vibrations have strong anharmonic character,
even at low temperatures. Several \textit{ab initio} studies of high
pressure phases of hydrogen, with and without anharmonic effects, have
been conducted recently to determine the stable structure of the high
pressure phases III and
IV\cite{pickard_density_2012,chen_room-temperature_2014,pickard_structure_2007,johnson_structure_2000}. In
particular, we use as test cases three structures from previous work,
labelled by their space group and number of atoms in the primitive
unit cell: \textit{Cmca}-4, \textit{Cmca}-12 and
\textit{C2/c}-24. These are molecular phases arranged in layers; the
\textit{C2/c}-24 phase is a candidate for phase III of
hydrogen\cite{pickard_structure_2007}. Previous work has shown that
anharmonicity has a significant effect on the vibrational energy of
these
structures\cite{borinaga_anharmonic_2016,monserrat_hexagonal_2016}.
The strongly anharmonic character of the nuclear vibrations in
hydrogen also implies that anharmonic effects that cannot be described
by simple 1-D mapping of the BO surface could be significant. This
provides an opportunity to test the ability of the VSCF+f method to
reduce the computational cost of calculations including the mapping of
2-D subspaces of the BO surface. The mapping of 2-D subspaces is
expensive in general, and therefore we have chosen the phase of solid
hydrogen with the fewest atoms in its unit cell, \textit{Cmca}-4, as a
test case for applying our improved method to mappings of 2-D subspaces
of the BO surface.

Lithium and zirconium both have body-centered cubic (\textit{bcc})
phases that are unstable at 0~K, and are stabilised at finite
temperatures. This property is shared with other elements such as
titanium and hafnium. In the case of Li, the structure at zero
temperature is a close-packed \textit{R9} structure, with the phase
transition to \textit{bcc} occurring at around
$70$~K\cite{hellman_lattice_2011,overhauser_crystal_1984,schaeffer_boundaries_2015}.
Zirconium, on the other hand, has an hexagonal \textit{hcp} structure
at room temperature, with the transition to \textit{bcc} occurring at
around $1366$~K\cite{grimvall_lattice_2012}. In both materials, the
\textit{bcc} phase is unstable at low temperatures due to the presence
of soft modes, which are dynamically stabilised at higher
temperatures.  Soft modes inherently have strong anharmonic character,
and an accurate description of anharmonicity is required to treat
these modes and the phase transition to the \textit{bcc} structure.
Several first principles vibrational studies including anharmonicity
have been reported for Li, Zr and other similar
elements\cite{antolin_fast_2012,grimvall_lattice_2012,hellman_lattice_2011,souvatzis_entropy_2008,petry_phonon_1991,heiming_phonon_1991,trampenau_phonon_1991}.
The well-known presence of significant anharmonic effects in the
\textit{bcc} phase of these materials makes them suitable test cases
for the VSCF+f method.

The rest of this work is organised as follows: In Sec.\ II we outline
the VSCF method and describe our implementation of it. In Sec.\ III we
describe the improved VSCF+f method and the testing of it, which is
the focus of the rest of this work. In Sec.\ IV we compare results
calculated using the VSCF+f and standard VSCF methods. Applications to
molecular and high pressure solid hydrogen are reported in Sec.\ IV.A,
mappings of 2-D subspaces of the BO surface in the \textit{Cmca}-4
hydrogen phase in Sec.\ IV.B, and the \textit{bcc} phases of Li and Zr
in Sec.\ IV.C. Finally, in Sec.\ V we summarise our results. All
equations are in Hartree atomic units, with
$\hbar=|e|=m_e=4\pi\varepsilon_0=1$.

\section{The VSCF method}

We work within the BO approximation in which the electronic and
nuclear motions are separated out, which leads to the vibrational
equation\cite{monserrat_anharmonic_2013}
\begin{equation} 
  \begin{aligned} \left( \sum_{p,\alpha}
      -\frac{1}{2m_\alpha} \nabla_{p\alpha}^2 +E_\text{el}(\mathbf{R})
    \right) \psi_\text{vib} (\mathbf{R}) = \\ \hat{H}_\text{vib}
    \psi_\text{vib} (\mathbf{R})= E_\text{vib} \psi_\text{vib}
    (\mathbf{R}). 
  \end{aligned}
\end{equation} where $p$ and $\alpha$ label
unit cells in the system and atoms within a unit cell respectively,
with $m_\alpha$ the mass of the $\alpha$\textsuperscript{th}
atom. $\mathbf{R}$ is a collective vector of all the nuclear
positions, $\psi_\text{vib}(\mathbf{R})$ is the vibrational
wavefunction, and $E_\text{vib}$ is the vibrational energy of the
system. The electronic energy for a given $\mathbf{R}$,
$E_\text{el}(\mathbf{R})$, acts as the potential in the vibrational
Hamiltonian $\hat{H}_\text{vib}$, and is generally known as the BO
surface\cite{monserrat_anharmonic_2013}.

Typically, this equation is solved approximately using the harmonic
approximation by expanding $E_\text{el}(\mathbf{R})$ up to quadratic
order in the displacement co-ordinates
$\mathbf{x}_{p\alpha}=\mathbf{r}_{p\alpha}-\mathbf{r}_{p\alpha}^0$.
Here, $\mathbf{r}_{p\alpha}$ and $\mathbf{r}_{p\alpha}^0$ are,
respectively, the displaced and equilibrium nuclear positions. We
calculate the harmonic potential by determining the matrix of force
constants
$\frac{\partial^2 E_\text{el} (\mathbf{R}^0)}{\partial
  x_{p\alpha;i} \partial x_{p'\alpha';j}}$,
using a finite differences method\cite{kunc_ab_1982}, and transforming
to reciprocal space to give the dynamical matrix:
\begin{equation} D_{i\alpha;j\alpha'}(\mathbf{q})=\frac{1}{N_p
    \sqrt{m_\alpha m_{\alpha'}}} \sum_{p,p'} \frac{\partial^2
    E_\text{el} (\mathbf{R}^0)}{\partial x_{p\alpha;i} \partial
    x_{p'\alpha';j}}
  e^{i\mathbf{q}\cdot(\mathbf{R}_p-\mathbf{R}_{p'})}.
\end{equation}
$\mathbf{R}_p$ is the position vector of the $p$\textsuperscript{th}
unit cell, $N_p$ is the number of unit cells in the system, and $i,j$
run over the Cartesian directions. The eigenvectors,
$w_{\mathbf{q}n;i\alpha}$, and eigenvalues, $\omega_{n\mathbf{q}}$, of
$D_{i\alpha;j\alpha'}(\mathbf{q})$ can then be found, with $n$
labelling the phonon branch index\cite{monserrat_anharmonic_2013}.

$\hat{H}_\text{vib}$ is then re-expressed in terms of harmonic normal
or phonon co-ordinates given by:
\begin{equation} u_{n\mathbf{q}} = \frac{1}{\sqrt{N_p}}
  \sum_{p,\alpha,i} \sqrt{m_\alpha} x_{p\alpha;i}
  e^{-i\mathbf{q}\cdot\mathbf{R}_p} w_{-\mathbf{q}n;i\alpha}.
\end{equation} 
Within the harmonic approximation, this gives a set of non-interacting
simple harmonic oscillators of frequencies $\omega_{n\mathbf{q}}$. In
the VSCF method, the principal axes
approximation\cite{jung_vibrational_1996} is used to include
anharmonic effects. Assuming that anharmonicity is a perturbation to
the harmonic approximation, and therefore a description of the BO
surface in terms of many 1-D subspaces is a good approximation, we
expand the BO surface as a series of $N$-D subspaces, using a basis
given by the harmonic normal modes, \cite{monserrat_anharmonic_2013}:
\begin{equation} \label{eq:BOExpansion}
  \begin{aligned} E_\text{el}
    (\mathbf{u}) =& E_\text{el}(\mathbf{0}) +\sum_{n,\mathbf{q}}
    V_{n\mathbf{q}} (u_{n\mathbf{q}}) +\\ &\frac{1}{2}
    \sum_{n,\mathbf{q}} \sum_{\substack{n'\mathbf{q}' \\ \neq
        n\mathbf{q}}} V_{n\mathbf{q};n'\mathbf{q}'}
    (u_{n\mathbf{q}},u_{n'\mathbf{q}'}) +
    \dotsb 
  \end{aligned}
\end{equation} 
$\mathbf{u}$ is a collective mapping amplitude vector. The
re-expression of $E_\text{el}(\mathbf{R})$ allows the mapping of the
BO surface along the directions corresponding to the harmonic phonons
and builds up an expression for $E_\text{el}(\mathbf{R})$, by means
that will be discussed below. With this expression in hand, the energy
can be minimised, using a Hartree product of 1-D states
$\Ket{\phi_{n\mathbf{q}}(u_{n\mathbf{q}})}$ as a trial
wavefunction. Solving the resulting VSCF equations gives a set of
anharmonic vibrational eigenstates, with their associated energy
eigenvalues and
wavefunctions\cite{monserrat_anharmonic_2013,bowman_self-consistent_1978}. A
perturbation theory can be constructed on these states, providing a
further correction to the energy, and a partition function can be
constructed, allowing the anharmonic free energy to be calculated at
any finite temperature\cite{monserrat_anharmonic_2013}.

\section{Improvements to the VSCF method}
In order to solve the VSCF equations we require the form of
$E_\text{el}(\mathbf{u})$. In previous work using this method, this is
typically obtained using DFT to calculate $E_\text{el}(\mathbf{u})$
for various values of $\mathbf{u}$ (i.e., various sets of atomic
positions), and then fitting a functional form to these DFT energy
results\cite{monserrat_electron-phonon_2014,engel_vibrational_2016,prentice_first-principles_2017}. 
To find the form of the 1-D terms $V_{n\mathbf{q}}(u_{n\mathbf{q}})$,
calculations are performed at several mapping points along the
direction given by the appropriate harmonic phonon; to find the form
of the 2-D terms
$V_{n\mathbf{q};n'\mathbf{q}'}(u_{n\mathbf{q}},u_{n'\mathbf{q}'})$,
calculations must be done at points on a grid, and so on. However,
these calculations rapidly become computationally expensive as we
include more terms in the expansion of the BO surface, and therefore
an increasingly large number of calculations is required. This
restricts us to including very few or none of the 2-D or higher
dimensional terms for all but the smallest systems.

Even when 2-D and higher terms are neglected, the mapping of the BO
surface is by far the most computationally expensive part of using the
VSCF method. In order to obtain an accurate fit to the BO surface, and
therefore an accurate solution to the VSCF equations, it is necessary
to converge the anharmonic correction to the energy,
$\Delta E_\text{anh}$, with respect to the number of mapping points
used per mapping direction, which introduces further computational
expense. Reducing the computational cost of the mapping is therefore
extremely desirable in order to speed up the calculations. One way to
reduce this cost would be to try and reduce the number of mapping
points required to reach convergence, by utilising information other
than the energy from the DFT calculations, such as the calculated
forces on the atoms.

The gradient of the BO surface at a given value of $\mathbf{u}$ can be
calculated from the forces on the atoms in the corresponding atomic
configuration
\begin{equation} \frac{\partial E_\text{el} (\mathbf{u})}{\partial
    u_{n\mathbf{q}}} = - \sum_{p\alpha;i} \frac{1}{\sqrt{N_p
      m_\alpha}}f_{p\alpha;i} e^{i\mathbf{q}\cdot\mathbf{R}_p}
  w_{\mathbf{q}n;i\alpha}. 
\end{equation} 
where $f_{p\alpha;i}$ is the force on the atom labelled by $p,\alpha$
in Cartesian direction $i$. Information about the forces can then be
fed into the fitting procedure, which then increases the accuracy of
the fit. In previous work, the DFT energy values have typically been
fitted using a cubic spline\cite{monserrat_anharmonic_2013}. In this
work we investigate using cubic and quintic splines in the fitting
procedure, to see whether either generally produces a more accurate
fit. Utilising a seventh order (heptic) spline was also tested, but
this approach suffered from overfitting and performed significantly
worse than the cubic and quintic splines. For this reason, heptic
splines will not be discussed further in this work.

Utilising forces to improve the mapping of the 1-D terms in the
expansion of the BO surface in equation \eqref{eq:BOExpansion} is a
simple extension of the usual splining procedure, but the task becomes
more complex when trying to improve the mapping of the higher
dimensional terms. The only higher dimensional terms considered here
are 2-D. A functional form for the BO surface is found by using a
series of 1-D splines along one direction, with the co-ordinate $u_1$,
and then using the results to fit a spline along the second direction,
with the co-ordinate $u_2$. Typically this would require knowledge of
the cross derivative
$\frac{\partial^2 E_\text{el}(\mathbf{u})}{\partial u_1 \partial u_2}$
at each sampling point\cite{press_numerical_1992}. We avoid this
requirement by using a simple cubic spline to obtain the form of the
function
$\left.\frac{\partial E_\text{el}(\mathbf{u})}{\partial
    u_2}\right|_{x_2} (u_1)$.
This represents the gradient along the direction defined by $u_2$ as a
function of $u_1$, for a fixed value of $u_2=x_2$. Obtaining this
function allows forces to be used in the fit along $u_2$. To ensure
the accuracy of the calculations including 2-D terms, we converge the
correction to the anharmonic vibrational energy due to these terms,
$\Delta E_\text{2-D} =
E_\text{2-D}^\text{anh}-E_\text{1-D}^\text{anh}$,
with respect to the number of mapping points used per direction.

Although including the force data in the fitting process should reduce
the number of calculations required to obtain convergence, the
accuracy of the forces themselves must be considered. If a variational
method is used to minimise the total energy in the DFT calculations,
the energy itself will be correct to second order errors in the charge
density. However, the error in the forces is instead linear with the
error in the charge density \cite{payne_iterative_1992}, and
calculations must be converged to within a strict tolerance to obtain
accurate forces. This requirement could potentially cancel out the
reduction in computational cost gained by reducing the number of
mapping points if the convergence tolerance is too strict, and our
results include tests to determine whether this is true or not. These
tests showed that this issue did not negatively affect the speed-up
obtainable with the VSCF+f method, with both methods breaking down at
the same level of energy convergence.

\section{Results}
All DFT calculations were performed using version 8.0 of the
plane-wave DFT code CASTEP\cite{clark_first_2005} and ultrasoft
pseudopotentials\cite{vanderbilt_soft_1990} generated
`on-the-fly'. Throughout, the ratio between the fast Fourier transform
(FFT) grid used for the electronic density and that used for the
Kohn-Sham states (the `grid scale' parameter in CASTEP) is $2.0$. The
local density approximation (LDA) was used for the
exchange-correlation functional in the hydrogen
calculations\cite{kohn_density_1996}, while the PBE functional was
used for lithium and zirconium\cite{perdew_generalized_1996}. Previous
work on solid hydrogen has shown that, while the exact quantitative
results of DFT calculations are strongly dependent on the choice of
functional, the qualitative results are similar for most
functionals\cite{azadi_fate_2013}, and that the LDA is a reasonable
choice for the purpose of tests on solid hydrogen. The PBE functional
has been used successfully in several previous studies of lithium,
zirconium and other similar
materials\cite{hellman_lattice_2011,souvatzis_entropy_2008}. Calculations
of the harmonic normal modes and energy were performed by determining
the matrix of force constants via a finite differences
method\cite{kunc_ab_1982}, before diagonalization of the dynamical
matrix. Atomic displacements of 0.01 bohr were used.

\subsection{Hydrogen}
As the lightest element, hydrogen is a good material in which to test
the VSCF+f method. Its vibrational motions explore the BO surface out
to large amplitudes due to its low mass, which introduces significant
anharmonic character. The method was first applied to molecular
hydrogen before moving onto solid hydrogen at 100~GPa. All hydrogen
calculations were performed at zero temperature and only the
zero-point energy was considered.

A plane-wave cut-off energy of 800~eV was used for the calculations on
molecular hydrogen, with a $5\times5\times5$ Monkhorst-Pack
$\mathbf{k}$-point grid\cite{monkhorst_special_1976}. As CASTEP uses
periodic boundary conditions, it is necessary to make the unit cell
large enough to prevent the molecule from interacting with its
periodic images. The frequencies of the harmonic phonon modes were
converged with respect to the size of the cubic unit cell, resulting
in a converged lattice constant of 8~\r{A}. The distance between the atoms
was allowed to relax, using the Broyden-Fletcher-Goldfarb-Shanno
(BFGS) method\cite{nocedal_numerical_2006} to converge the forces on
the atoms to within $0.001$~eV~$\text{\r{A}}^{-1}$, before
single-point energy calculations were conducted.

\begin{figure}
\centering
\includegraphics[width=0.49\textwidth]{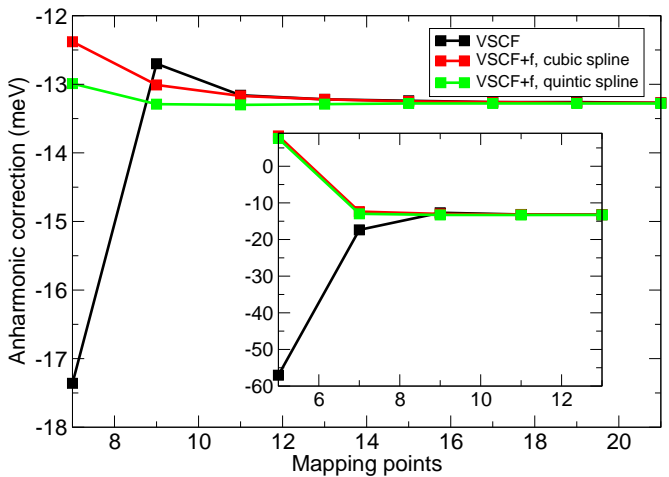}
\caption{Convergence of anharmonic correction to the energy at 0~K for
  H$_2$, $\Delta E_\text{anh} = E_\text{anh}-E_\text{har}$, with
  respect to the number of mapping points used per mapping direction,
  for the basic VSCF method, as well as the VSCF+f method, fitting
  with both cubic and quintic splines. The inset shows the results at
  low numbers of mapping points on a different energy scale.}
\label{fig:H2Convergence}
\end{figure}

Once the harmonic calculations were completed, a range of different
numbers of mapping points per direction from 7 to 27 were used to map
the BO surface of the hydrogen molecule. The performance of the basic
VSCF method was compared to the VSCF+f method using both cubic and
quintic splines in the fitting process. This was repeated with the
energy convergence tolerance of the calculations set to
$10^{-10}, 10^{-6}, 10^{-4}$ and $10^{-2}$~eV per SCF cycle, to test
whether this affected the accuracy of the forces, and therefore the
accuracy of the VSCF+f fitting relative to the normal fitting
procedure. Fig.\ \ref{fig:H2Convergence} shows the convergence of the
anharmonic correction to the zero point energy,
$\Delta E_\text{anh} = E_\text{anh}-E_\text{har}$, with respect to the
number of mapping points for the three different methods. Here the
energy convergence tolerance was set to $10^{-6}$~eV per SCF cycle. It
can be seen that including the forces in the fitting process
significantly improves the convergence of $\Delta E_\text{anh}$, with
the quintic spline fit performing even better than the cubic
spline. This suggests that including forces in the fitting process can
significantly improve the efficiency of the VSCF method, and that
utilising a quintic spline allows fitting of the BO surface even more
accurately for the same number of DFT calculations, especially for low
numbers of mapping points. An almost identical set of results was
found for energy convergence tolerances from $10^{-4}$~eV up to
$10^{-10}$~eV per SCF cycle, with the VSCF+f method converging more
rapidly with the number of mapping points. For an energy convergence
tolerance of $10^{-2}$~eV per SCF cycle, both the VSCF and VSCF+f
methods failed to converge with less than 27 mapping points per
direction. This shows that for a range of energy convergence
tolerances, the VSCF+f fitting method is still able to outperform the
basic VSCF method and map the BO surface accurately at a lower
computational cost. Using a quintic spline improves the quality of the
fit still further.

\begin{figure*}
\centering
\begin{subfigure}{0.3\textwidth}
\includegraphics[width=\textwidth]{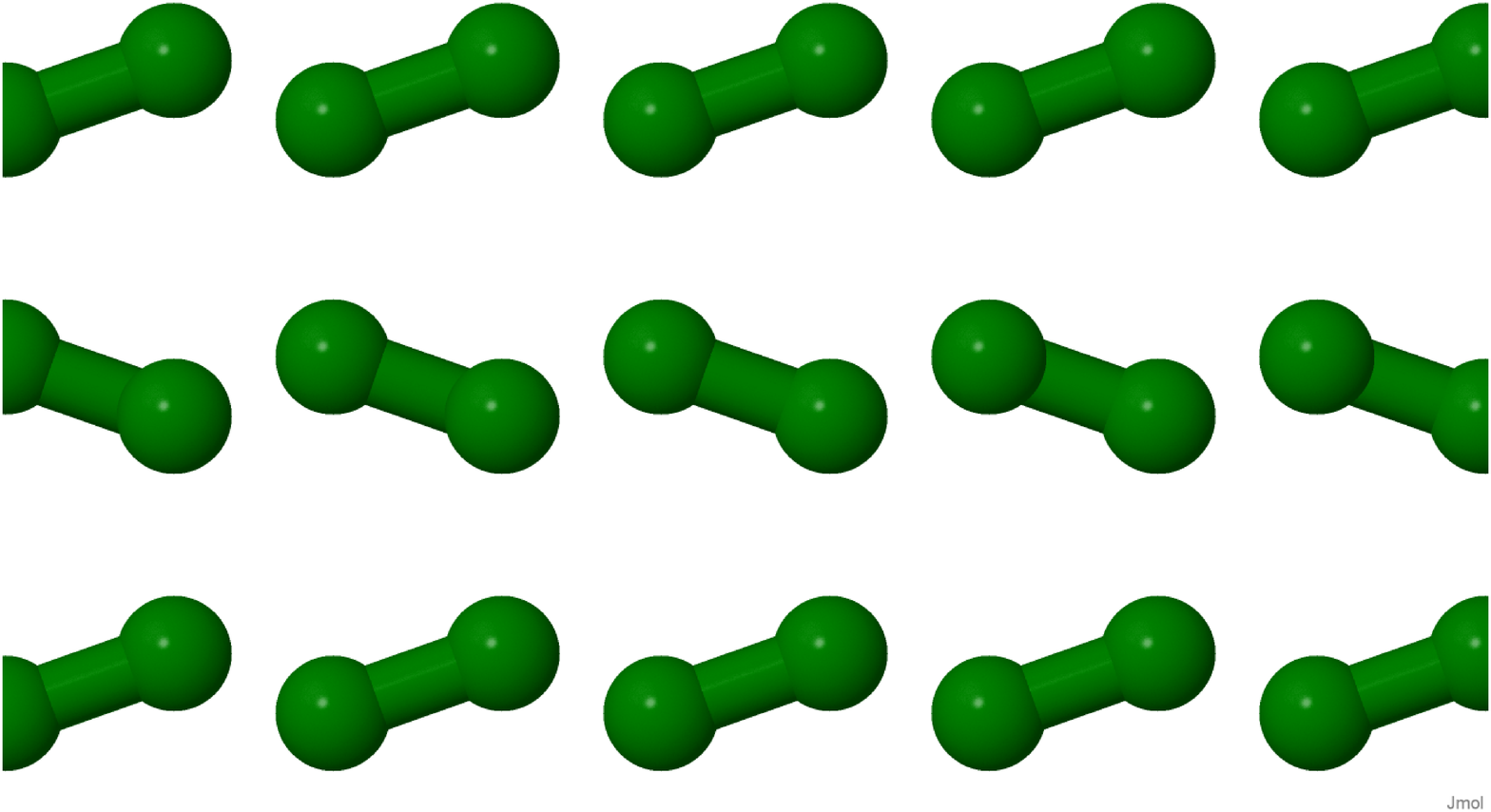}
\caption{\emph{Cmca}-4}
\end{subfigure}
~
\begin{subfigure}{0.3\textwidth}
\includegraphics[trim={4cm 2.75cm 3.75cm 3.5cm},clip,width=\textwidth]{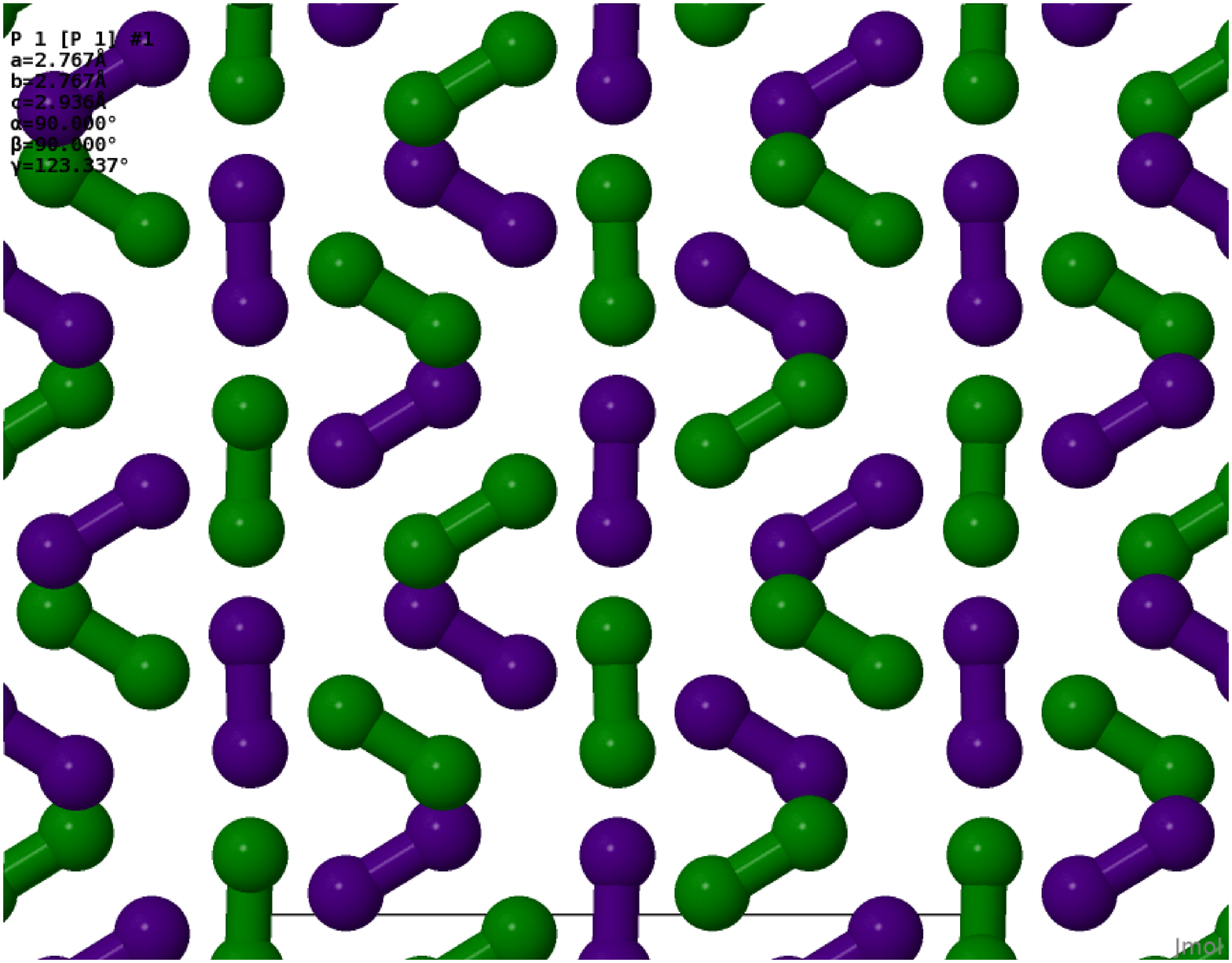}
\caption{\emph{Cmca}-12}
\end{subfigure}
~
\begin{subfigure}{0.3\textwidth}
\includegraphics[trim={4cm 1cm 3.75cm 2cm},clip,width=\textwidth]{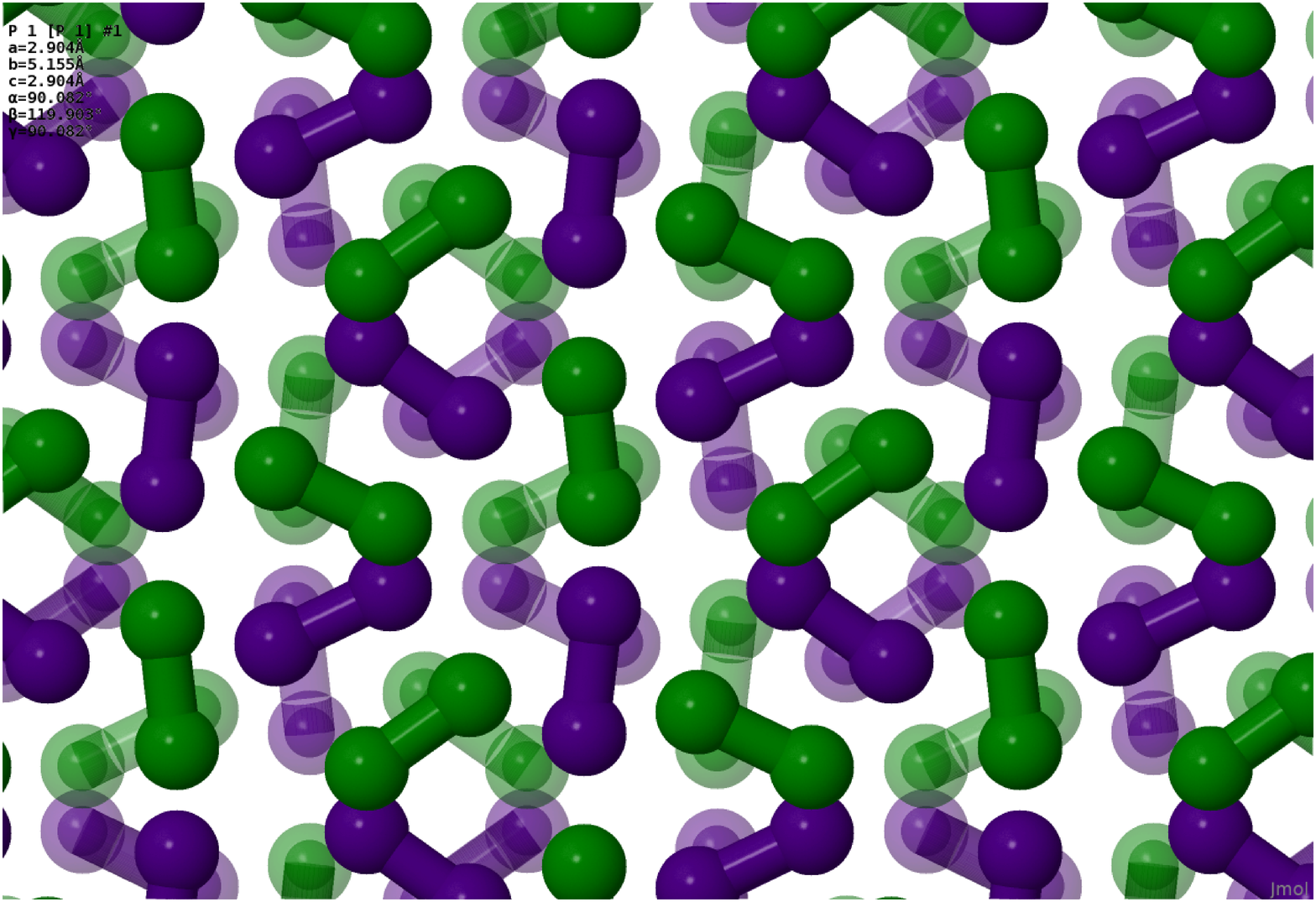}
\caption{\emph{C2/c}-24}
\end{subfigure}
\caption{Views of the structures of solid hydrogen considered at
  100~GPa. Both of the \emph{Cmca} structures are shown looking along
  the $x$-axis, while the \emph{C2/c}-24 structure is shown looking
  along the $y$-axis. As these are layered structures, atoms in
  inequivalent layers are denoted by different colours. Green, purple,
  translucent green and translucent purple denote the first, second,
  third and fourth inequivalent layers, respectively.}
\label{fig:SolidHStructures}
\end{figure*}

\begin{figure}
\centering
\includegraphics[width=0.49\textwidth]{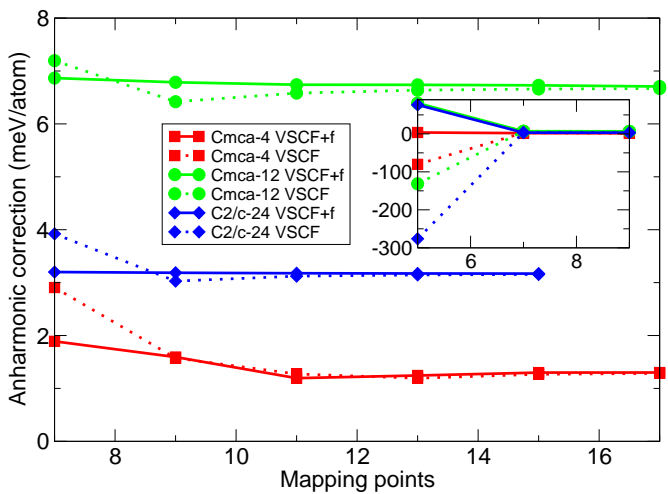}
\caption{Convergence of the anharmonic correction to the energy at
  0~K, $\Delta E_\text{anh}$, of the \emph{Cmca}-4, \emph{Cmca}-12 and
  \emph{C2/c}-24 phases of solid hydrogen at 100~GPa, with respect to
  the number of mapping points used per mapping direction. The
  convergence of the basic VSCF method as well as the VSCF+f method of
  fitting with a quintic spline are shown. The inset shows the results
  at low numbers of mapping points on a different energy scale.}
\label{fig:HSolidConvergence}
\end{figure}

With the results from the hydrogen molecule in mind, we turn our
attention to the case of high pressure solid hydrogen. Three different
structures of solid hydrogen were considered at a pressure of 100~GPa
- \textit{Cmca}-4, \textit{Cmca}-12 and \textit{C2/c}-24, as
previously described. Fig.\ \ref{fig:SolidHStructures} gives a view of
these three structures, showing that they are molecular in
nature. These structures have all been studied previously with DFT
over a range of pressures\cite{pickard_density_2012}. Again, once
harmonic calculations were completed, the covergence of the anharmonic
correction to the zero point energy per atom with respect to the
number of mapping points was calculated for a single unit cell of each
structure. An plane-wave cut-off energy of 1000~eV and an energy
convergence tolerance of $10^{-6}$~eV per SCF cycle was used
throughout, with Monkhorst-Pack grids of size $28\times28\times16$,
$18\times18\times4$ and $16\times8\times16$ for the \emph{Cmca}-4,
\emph{Cmca}-12 and \emph{C2/c}-24 structures respectively. Fig.\
\ref{fig:HSolidConvergence} shows the convergence of
$\Delta E_\text{anh}$ per atom for all three structures for the basic
VSCF and the improved VSCF+f quintic spline methods. Again,
convergence was reached with fewer numbers of mapping points per
direction using the VSCF+f method than with the VSCF method,
especially for the \emph{Cmca}-12 and \emph{C2/c}-24 structures. In
the latter case, including forces in the fitting reduces the
computational cost by around 40\%. This further implies that the
VSCF+f method is robust and improves on the efficiency of the basic
VSCF method.

\subsection{Two-dimensional subspaces}
A natural next test of the VSCF+f method is to apply it to mapping 2-D
subspaces of the BO surface. This poses a more significant challenge
than the 1-D terms considered up to now, as interpolating data in two
dimensions is required.

Mapping the BO surface in two dimensions is expensive, and can only
feasibly be done for small systems. Here, we focus on the
\emph{Cmca}-4 structure of solid hydrogen discussed previously, which
possesses twelve potential mapping directions, corresponding to the
twelve harmonic phonon modes, labelled with numbers from 1 to
12. Three of these modes are acoustic, and thus have zero
frequency. To minimise the computational cost further, we consider
only two of the many 2-D subspaces in this system -- those
corresponding to the directions described by the harmonic modes 4 and
5, and 4 and 7. The harmonic frequencies of modes 4, 5, and 7 are
$69.4$, $74.0$ and $114$~meV, respectively, and the displacement
patterns corresponding to each of these mapping directions can be
found in the Supplemental Material. These subspaces were chosen by
conducting a preliminary mapping of all 2-D subspaces with a low
number of mapping points, and taking only those with significant
corrections to the 1-D description of the BO surface. The subspaces
where the mapping entered parts of energy minima corresponding to
structures significantly lower in energy than the \emph{Cmca}-4
structure were also neglected. The two subspaces presented here were
then chosen as being representative of those remaining. The same
cut-off energy and Monkhorst-Pack grid was used as in the calculations
of Sec.\ IV.A, but an energy convergence tolerance of $10^{-10}$~eV
was used to ensure accurate forces.

\begin{figure*}
\centering
\begin{subfigure}{0.85\textwidth}
\includegraphics[trim={0.75cm 0cm 1cm
  3cm},clip,width=0.51\textwidth]{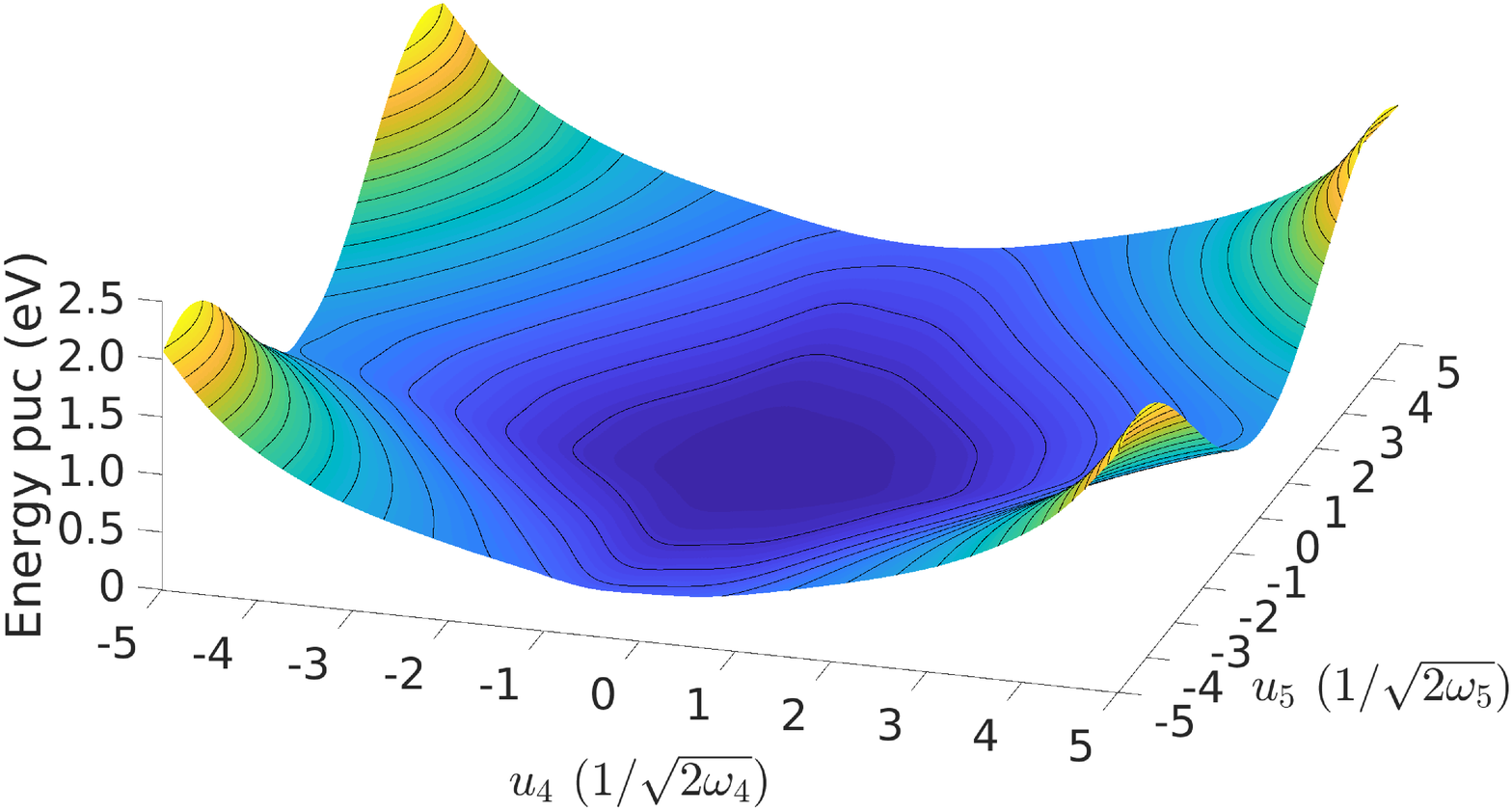}
\includegraphics[width=0.47\textwidth]{Fig4aii.eps}
\caption{Subspace corresponding to directions 4 and 5}
\end{subfigure}
~
\begin{subfigure}{0.85\textwidth}
\includegraphics[trim={1cm 0cm 0cm
  2cm},clip,width=0.51\textwidth]{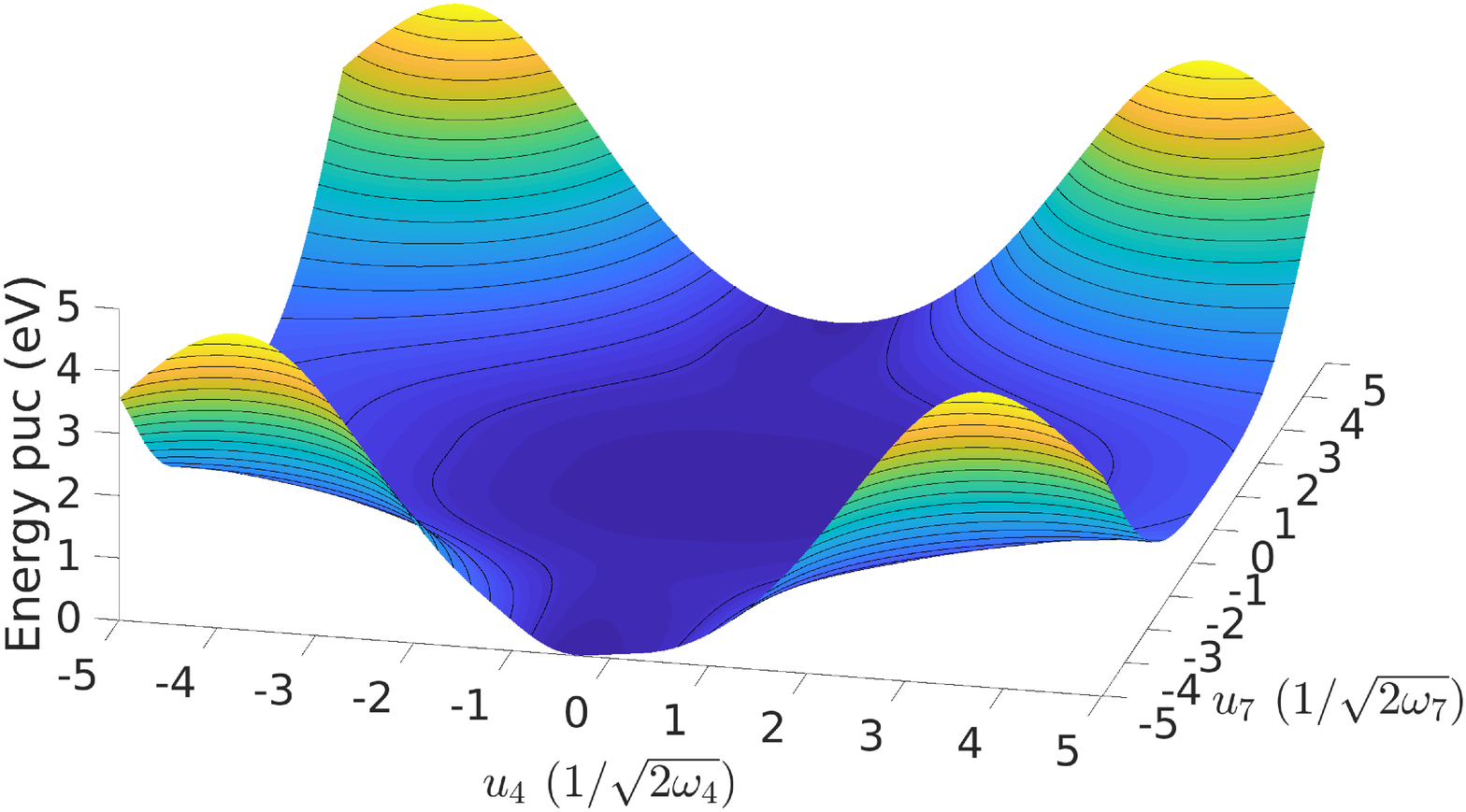}
\includegraphics[width=0.47\textwidth]{Fig4bii.eps}
\caption{Subspace corresponding to directions 4 and 7}
\end{subfigure}
\caption{Results of anharmonic vibrational calculations for the
  \emph{Cmca}-4 structure of solid hydrogen including selected 2-D
  subspaces of the BO surface. The left-hand column shows BO surfaces
  mapped in the labelled subspace, where 'puc' stands for 'per unit
  cell', while the right-hand column shows the convergence of the
  correction to the vibrational energy due to the relevant 2-D term
  with respect to the number of mapping points used.}
\label{fig:HSolidCouplingConvergence}
\end{figure*}

Fig.\ \ref{fig:HSolidCouplingConvergence} shows the results of tests
including the mapping of 2-D subspaces in the \emph{Cmca}-4 solid
hydrogen structure. The two rows of figures correspond to the two
subspaces mapped. The left-hand column shows the BO surface mapped in
the relevant subspace, and the right-hand column shows the convergence
of the correction to the energy due to 2-D terms $\Delta E_\text{2-D}$
with respect to the number of mapping points used per mapping
direction. All energies were again calculated at zero temperature. The
convergence graphs show that utilising forces in the mapping of the BO
surface in two dimensions improves the results relative to the
converged final value, especially for small numbers of mapping points,
although the improvement is not as pronounced as in the 1-D case. This
could be due to the small size of the energy scales in question -- the
energies shown are all smaller than 1~meV per atom, which is around
the finest energy scale that such anharmonic calculations can
reasonably be assumed to be accurate to. The small size of the
corrections due to 2-D terms compared to those seen for 1-D terms
shows that the neglect of such higher-order terms in the BO surface
expansion of equation \eqref{eq:BOExpansion} is justified.  The
ability of the VSCF+f method to show improvement even at such small
energy scales again demonstrates its capabilities, even in cases
including mapping of 2-D subspaces of the BO surface.

\subsection{Lithium and zirconium}
Finally, we consider two systems in a very different regime from
molecular or solid hydrogen - the metals lithium and zirconium, in
their \textit{bcc} phases. The main anharmonic contributions to the
vibrational energy in these systems arise from soft modes, which
necessarily have some quartic character. Here, we focus our attention
on mapping the BO surface along the direction defined by these soft
modes, and how utilising the VSCF+f procedure can improve this
mapping. In addition to calculations at zero temperature, the effect
of anharmonicity on the free energy at finite temperature is also
considered, as the \textit{bcc} phase only becomes stable
experimentally at $70$~K in Li and $1366$~K in Zr. Thermal expansion
effects are not included in these calculations.

For the calculations in lithium and zirconium, we used the recently
introduced non-diagonal supercells method to reduce the computational
cost of sampling the vibrational Brillouin
zone\cite{lloyd-williams_lattice_2015}. Typically, diagonal supercells
are used to sample the vibrational BZ, with an $N\times N\times N$
sampling grid requiring an $N\times N\times N$ supercell containing
$N^3$ unit cells. The non-diagonal supercells method allows such an
$N\times N\times N$ sampling to be done using a series of non-diagonal
supercells each containing a maximum of $N$ unit cells, significantly
reducing the computational cost. However, as the aim of this work is
to consider methods for improved fitting of the BO surface, rather
than to conduct high-accuracy calculations on these well-studied
materials, we did not attempt to completely converge our results with
respect to the sampling of the vibrational BZ. Even with the
non-diagonal supercell method the computational cost increases rapidly
with increasing sampling grid size. Instead, we used a
$8\times8\times8$ sampling of the vibrational BZ as a compromise
between accuracy and speed. 

As lithium and zirconium are metallic, it is necessary to use partial
band occupancies to eliminate discontinuities in the energy during the
SCF minimisation. This is done by artificially giving the Kohn-Sham
quasiparticles a finite temperature to smear out their energy
levels. In these calculations, a smearing width of $0.2$~eV,
corresponding to $2320$~K, is used. The calculated DFT energies do
depend on the size of this smearing, but the effect on the overall
shape of the BO surface is small, meaning the results of the
vibrational calculations are largely unaffected by the choice of
smearing width. The Monkhorst-Pack grids in all supercells used had a
spacing of $0.025$~$\text{\r{A}}^{-1}$, corresponding to a
$16\times16\times16$ grid in the unit cell. An energy cut-off of
$1500$~eV was used in all calculations. To obtain these parameter
values, the harmonic vibrational energy was converged to within
$1$~meV with respect to the Monkhorst-Pack grid size and energy
cut-off. This constitutes a somewhat stronger convergence criterion
than that used in other work\cite{lejaeghere_reproducibility_2016}, as
it is important to ensure the harmonic results are accurate before
using them as a basis for further calculations.

Our results for Li show that any differences in the anharmonic
vibrational energy arising from the three different fitting methods
are negligible, down to scales of 0.1~meV. Because of this we do not
present the results for Li here, although we do discuss them. The
negligible difference between fitting methods arises from the fact
that, apart from along the direction defined by the soft modes
present, the vibrational properties of Li are described well by the
harmonic approximation, despite its low
mass\cite{taole_anharmonic_1978}. This means that a very good fit to
the BO surface along most modes can be found with small numbers of
mapping points (as in the standard finite displacement method for
calculating harmonic frequencies), and so all three methods agree very
well. Even in the case of the soft modes, which must necessarily
contain some anharmonic character, the double-well structure is not
very pronounced, with the overall BO surface appearing essentially
quadratic. To see this double-well structure, where the two minima are
very close to the central maximum, a finer than usual mapping of the
BO surface proved necessary. Our calculations show these double wells
are quite shallow, meaning that even at zero temperature the BO
surface looks essentially harmonic. Our results imply that the
\textit{bcc} structure of Li is dynamically stable at zero
temperature, although experimental results show that the \textit{bcc}
phase becomes stable above $70$~K. This disagreement could potentially
be caused by incomplete convergence with respect to sampling of the
vibrational BZ, or higher order terms in the expansion of the BO surface
of equation \eqref{eq:BOExpansion}, as
well as by the errors inherent in DFT. A different
exchange-correlation functional might give results closer to
experiment.

\begin{figure}
\centering
\includegraphics[width=0.49\textwidth]{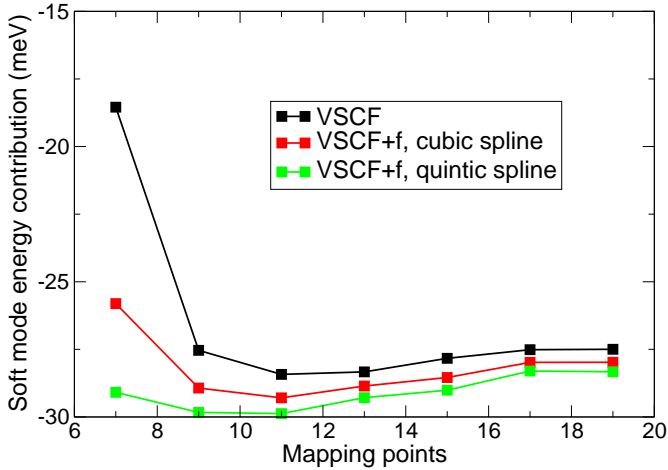}
\caption{Convergence with respect to the number of mapping points used
  per mapping direction of  contribution of 1-D terms corresponding to the mapping
  directions defined by soft modes to the total anharmonic vibrational
  energy of the \textit{bcc} phase of zirconium at $0$~K. The
  convergence of the basic VSCF method as well as the VSCF+f method
  using both a cubic spline and a quintic spline are shown.}
\label{fig:ZrAnhSoftModes}
\end{figure}

\begin{figure}
\centering
\includegraphics[width=0.49\textwidth]{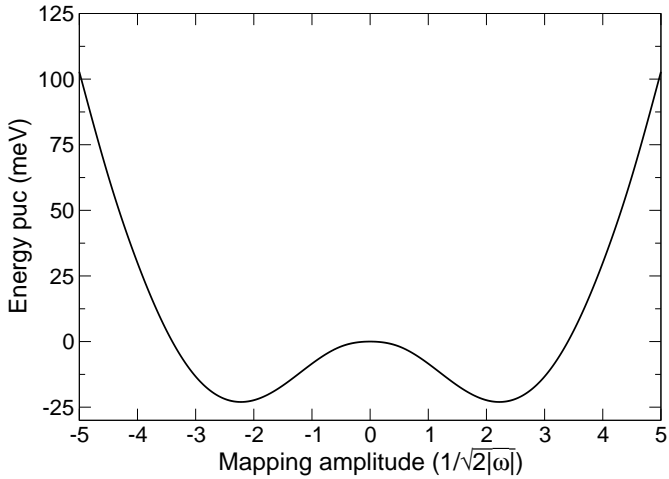}
\caption{The shape of the BO surface as mapped along one of the soft
  modes of zirconium; 'puc' stands for 'per unit cell'.}
\label{fig:ZrSoftMode}
\end{figure}

\begin{figure}[t]
\centering
\includegraphics[width=0.49\textwidth]{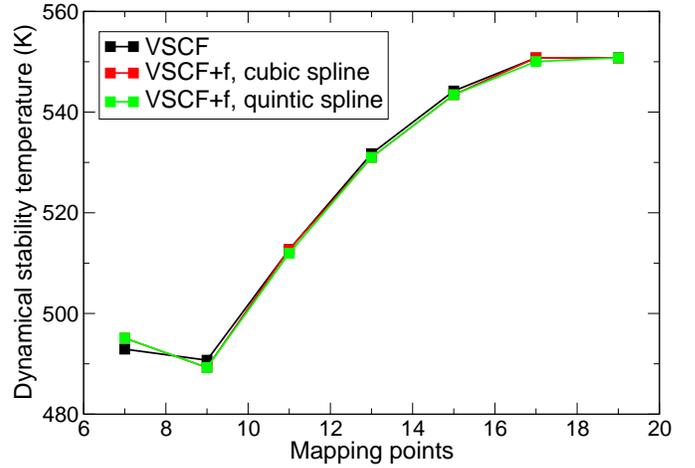}
\caption{Convergence of the temperature at which the \textit{bcc}
  phase of zirconium becomes dynamically stable with respect to the
  number of mapping points used per mapping direction. Results using the basic VSCF
  method and the VSCF+f method using both a cubic spline and a quintic
  spline are shown. The VSCF+f cubic spline results are hidden behind
  the VSCF+f quintic spline results.}
\label{fig:ZrTemp}
\end{figure}

The results for zirconium tell a different story to those of
lithium. Fig.\ \ref{fig:ZrAnhSoftModes} shows the variation with the
number of mapping points of the sum of the lowest eigenvalues
calculated for each of the 1-D terms resulting from mapping the BO
surface along a soft mode. The differences between the values obtained
by the different fitting methods for the sum of the lowest eigenvalues
of all the 1-D terms (which would represent the zero-point energy if
the \textit{bcc} structure of Zr was dynamically stable at 0~K) are
small, as in Li. However, there are much more significant differences
in the contribution of the soft modes to the anharmonic energy, as
seen in Fig.\ \ref{fig:ZrAnhSoftModes}. Although the effect is much
less pronounced than that seen for hydrogen in Fig.\
\ref{fig:HSolidConvergence}, it is clear that, for low numbers of
mapping points, including force data improves the fit to the BO
surface. It is also evident that the quintic spline gives a better fit
than the cubic spline when the forces are used. This demonstrates that
the VSCF+f method can improve on the basic VSCF method in this type of
system, as well as the hydrogen systems explored previously.

The soft modes in Zr are more numerous, and mapping the BO surface
along the directions defined by them gives much more pronounced
double-well structures than in Li, meaning that the structure is not
dynamically stable at zero temperature. An example of the pronounced
double-well structure of the BO surface mapped along one of the soft
modes in Zr is shown in Fig.\ \ref{fig:ZrSoftMode}. Our results can be
used to calculate the temperature at which the \textit{bcc} phase is
stabilised dynamically, by calculating the internal energy at a range
of temperatures and finding where it becomes positive. The results are
shown in Fig.\ \ref{fig:ZrTemp} for a range of mapping points and the
three fitting methods used. At these higher temperatures, it can be
seen that the differences between the three fitting methods, visible
at zero temperature in Fig.\ \ref{fig:ZrAnhSoftModes}, are much less
significant -- the differences are mostly washed out by the overall
vibrational energy increasing. Our calculations predict that the
\textit{bcc} structure of Zr should become dynamically stable above
about $520$~K, which is significantly lower than the observed
transition temperature of $1366$~K. This disagreement could again
potentially be caused by the incomplete convergence with respect to
the vibrational BZ sampling, higher-order terms in the BO surface
expansion, thermal expansion effects, or errors inherent in DFT
itself.

\section{Conclusions and future work}

In summary, we have shown that the efficiency of the vibrational
self-consistent field method proposed in Ref.\
\onlinecite{monserrat_anharmonic_2013} can be significantly improved
by using both energy and force data from DFT calculations when mapping
the BO energy surface.  Tests of this method on molecular and
high-pressure solid hydrogen, including the contribution of 2-D
subspaces of the BO surface to the energy, as well as on lithium and
zirconium in the \textit{bcc} structure, show that the VSCF+f method
agrees well with the basic VSCF method, but significantly reduces the
computational cost involved.

Our results show that the VSCF+f method allows us to perform accurate
calculations of anharmonic corrections to the harmonic phonon model
for a significantly reduced computational cost compared to the basic
VSCF method. The next step will be to apply the VSCF+f method to new
systems in which large anharmonicities are of interest.

\section{Acknowledgements}

The authors would like to thank Bartomeu Monserrat for useful
discussions and comments on the manuscript. Calculations were
performed on the Cambridge High Performance Computing Service facility
and the ARCHER facility of the UK National Supercomputing Service.
R.J.N.\ acknowledges financial support from the Engineering and
Physical Sciences Research Council (EPSRC) of the U.K.\ under Grant
No.\ EP/J017639/1.  Computational resources were provided by the
Archer facility of the U.K.'s national high-performance computing
service (for which access was obtained via the UKCP consortium, EPSRC
Grant No.\ EP/K014560/1). Supporting research data may be freely
accessed at https://doi.org/XX.XXXXX/XXX.XXXX, in compliance with the
applicable Open Data policies.

\bibliography{ForcesBib}

\end{document}


\title{Supplementary Information: Using forces to accelerate first-principles anharmonic vibrational calculations}

\author{Joseph C.\ A.\ Prentice \and R.\ J.\ Needs}

\date{\today}

\maketitle

\section{Pseudopotentials}

All density functional theory calculations were performed using {\sc CASTEP} version $8.0$, and its own ``on-the-fly'' ultrasoft pseudopotentials. The definition strings for the pseudopotentials were:

\begin{itemize}
\item H: \begin{verbatim}1|0.6|1|6|10|10(qc=8)\end{verbatim}
\item Li: \begin{verbatim}1|1.0|14|16|18|10U:20(qc=7)\end{verbatim}
\item Zr: \begin{verbatim}3|2.1|7|8|9|40U:50:41:42\end{verbatim}
\end{itemize}

\section{Equilibrium unit cell configurations}

The unit cells for the structures used in this work, containing the atoms at their equilibrium positions, are given in the \texttt{.cif} files \texttt{H2.cif}, \texttt{cmca4.cif}, \texttt{cmca12.cif}, \texttt{c2c24.cif}, \texttt{Li.cif} and \texttt{Zr.cif}.

\section{Harmonic mode displacement patterns}

The displacement patterns corresponding to the mapping directions used in the mapping of 2-D subspaces of the BO surface of the \textit{Cmca}-4 structure of solid hydrogen are given below. They correspond to the displacement patterns of harmonic modes with frequencies of $69.4$, $74.0$ and $114$~meV, labelled as 4, 5 and 7 respectively. Each row shows the displacement of a H atom in the three Cartesian directions, in the same order as the atoms are listed in the file \texttt{cmca4.cif}.

\begin{itemize}
\item Direction 4:
\begin{center}
\begin{verbatim}
  x     y     z
  0.00  0.75 -1.00
  0.00  0.75  1.00
  0.00 -0.75  1.00
  0.00 -0.75 -1.00
\end{verbatim}
\end{center}

\item Direction 5:
\begin{center}
\begin{verbatim}
  x  y  z
  0  1  0
  0 -1  0
  0  1  0
  0 -1  0
\end{verbatim}
\end{center}

\item Direction 7:
\begin{center}
\begin{verbatim}
  x  y  z
  1  0  0
 -1  0  0
 -1  0  0
  1  0  0
\end{verbatim}
\end{center}

\end{itemize}